\documentclass[12pt]{article}
\usepackage{amsmath}
\usepackage{graphicx,psfrag,epsf}
\usepackage{enumerate,color}
\usepackage{todonotes}
\usepackage{natbib}
\usepackage{url} 

\newcommand{\blind}{1}

\usepackage{amsthm}
\usepackage{graphicx,hyperref}
\usepackage{amssymb}
\usepackage{amsmath}
\usepackage{multirow}
\usepackage{float}
\usepackage{mathtools}
\mathtoolsset{showonlyrefs=true}

\usepackage{subfig}
\usepackage{graphicx}
\usepackage{epsfig}
\usepackage{epstopdf}
\usepackage{amsthm}

\date{}
\begin{document}

\def\spacingset#1{\renewcommand{\baselinestretch}%
{#1}\small\normalsize} \spacingset{1}


\if1\blind
{
  \title{\bf The Joint Projected and Skew Normal}
    \author{Gianluca Mastrantonio \hspace{.2cm}\\
    Department of Mathematical Science, Polytechnic of Turin, Turin
    }
  \maketitle
} \fi

\if0\blind
{
  \bigskip
  \bigskip
  \bigskip
  \begin{center}
    {\LARGE\bf On initial direction, orientation and discreteness in the analysis of circular variables}
\end{center}
  \medskip
} \fi
\begin{abstract}

We introduce a new multivariate circular linear distribution suitable for modeling direction and speed in (multiple) animal movement data. To properly account for specific data features, such as heterogeneity and time dependence, a hidden Markov model is used. Parameters are estimated under a Bayesian framework and we provide computational details to implement the Markov chain Monte Carlo algorithm.

The proposed model is applied to a dataset of six free-ranging
Maremma Sheepdogs. Its predictive performance, as well as the interpretability of the results, are compared to those given by hidden Markov models
built on
all the combinations of von Mises (circular), wrapped Cauchy (circular), gamma (linear) and Weibull (linear) distributions
\end{abstract}
\noindent%
{\it Keywords:} Animal movement,
Circular-linear distribution,
Multivariate projected normal,
Multivariate skew normal,
Dirichlet Process,
Hidden Markov model
\vfill

\newpage

\spacingset{1.2} 


\section{Introduction}

In the recent literature, the interest in modelling animal movement data, with the goal to understand the animals behaviour, is increasing. Animal movement modeling has a long history,  dating back to the diffusion model   of \cite{Brownlee1912}. 
A wide range of different models  have been proposed, such as stochastic differential equations   \citep{Blackwell2003},   mixture of random walks \citep{morales2004},   Brownian bridge \citep{Horne2007},   agent-based model \citep{Hooten2010},  mechanistic approach  \citep{McClintock2012} or  the  continuous-time discrete-space model  \citep{hanks2015}.

Animal movement data often  take the form of a bivariate time series of spatial coordinates    obtained by equipping an   animal  with a tracking device, e.g. a GPS collar, that  records locations at different times. These type of data are called \emph{telemetry data}.  From  telemetry data,  movements are measured by computing  the so-called \emph{movement metrics} \citep{Patterson2008}, such as  \emph{step-length} and \emph{turning-angle}, see for example \cite{DElia2001},  \cite{Jonsen2005} or \cite{Ciucci2009}.  Observed metrics are random variables and, accordingly, a parametric distribution is often needed to model these data.

Improved communication systems, shrinking battery sizes  and  the prices drop of GPS devices, have  led to an increasing availability  of datasets \citep{Cagnacci2010}. 
The  data,  often  freely available (see for example the \emph{movebank data repository} at  \url{www.movebank.org}), have a complex structure  because the animal behaviour changes over time and the occurrences of  behavioural modes are not time-independent \citep{Houston1999} (temporal dependence) \citep{morales2004}, each behaviour  is characterized by different distributions of the associated movement metrics (heterogeneity) and there is dependence in the movement metrics between and within animals (multivariate associations) \citep{langrock2014b}.

%
%

Time dependence and heterogeneity have been addressed,  in the literature,   by using 
%
hidden Markov models (HMMs), see for example  \cite{Franke2004},  \cite{Holzmann2006},  \cite{Jonsen:2007}, \cite{Eckert2008}, \cite{JANE2009}, \cite{Schliehe2012} or  \cite{langrock2014b}.
In most of the animal movement applications, the number of behavioural modes is fixed at priori using external knowledge.

The multivariate interactions between animals have been modeled in several ways. \cite{JANE2006}  assume a common distribution for some individual-level parameters that allows inference about population-level parameters.  
\cite{langrock2014b} propose a parent-child structure, where the animals (the children) are all attracted to an abstract point (the parent) while in   \cite{Morales2010}  the animals   are treated as independent assuming that the movement of one animal  is  representative of the group's overall movement.

A natural way to model the multivariate interactions is  to define a  joint distribution for the movement metrics that, generally,  are composed by   measures of speed (e.g. the  step-length) and  direction (e.g the turning-angle). The direction is a   \emph{circular variable}, it represents an angle or a point over a circumference and due to the particular topology of the circle must be treated differently from the linear (or inline) ones, e.g. variables defined over $\mathbb{R}$ or $\mathbb{R}^+$; for an introduction on  circular variables  see the book of  \cite{Merdia1999} or \cite{Jammalamadaka2001}.  A joint modelling of  step-lengths and turning-angles requires a multivariate distribution for circular-linear variables but in the literature have been proposed only distribution for cylindrical data \citep{SenGupta2004,Sengupta2014,mastrantonio2015,Abe2015}, i.e.  one circular and one linear variable.

In this work we are interested in finding the behavioural modes of  six free-ranging sheepdogs attending livestock  \citep{vanbommel2014,vanbommel20142} and understand  how they interact. Motivated by the data at hand we introduce   a new  flexible multivariate circular-linear distribution with dependent components, called the \emph{joint projected  and skew normal},   based on the skew normal of \cite{sahu2003} and on a multivariate extension of the projected normal \citep{Wang2013}. This distribution allows us to model jointly the movement metrics,  introducing dependence among animals. 
The proposal is used as emission distribution in an HMM. We propose to estimate the parameters in a non-parametric Bayesian framework, relying on the \emph{sticky hierarchical Dirichlet process-hidden Markov model} (sHDP-HMM) of \cite{fox2011}. This allows to jointly estimate  model parameters and  the number of behavioural modes without fixing  it a priori. We show how to estimate the parameters using a Markov chain Monte Carlo (MCMC) algorithm. As a by-product,  our MCMC implementation  solves the well-known identification problem  of the univariate projected normal distribution \citep{Wang2013}.


The paper is organized as follows. In Section \ref{sec:joints} we introduce the proposed distribution and in Section \ref{sec:idsolve}  we show how to estimate its parameters in a Bayesian framework. In Section \ref{sec:HPD-HMM} we introduce the HMM (Section \ref{sec:HMM}) and the non-parametric extension (Section \ref{sec:stick1}). 
In Section \ref{sec:real} we apply the model to the real data example  and in Section \ref{sec:conf}  we compare the proposed emission distribution with the  most used in the literature.
The paper ends with some conclusion remarks in Section \ref{sec:conc}.


\section{The multivariate circular-linear  distribution} \label{sec:joints}

In this Section we  introduce the projected normal, its multivariate extension and the skew normal of \cite{sahu2003}  used to built the  new multivariate circular-linear distribution.



\subsection{The multivariate   projected normal distribution} \label{sec:multi} \label{sec:multi2}


Let $\mathbf{W}_i = (W_{i1},W_{i2})^{\prime}$  be a $2$-dimensional random variable, normally distributed with mean vector  $\boldsymbol{\mu}_{wi}$ and covariance matrix $\boldsymbol{\Sigma}_{wi}$.  The random variable
\begin{equation} \label{eq:tranpn}
	\Theta_i = \text{ atan}^* \frac{W_{i2}}{W_{i1}} \in [0 ,2 \pi),\footnote{$\text{ atan}^*$ is a modified  arctangent function defined  in \cite{Jammalamadaka2001} pag. 13.}
\end{equation}
is a circular variable, i.e. a variable that represents an angle  over the unit circle,  distributed as a \emph{projected normal} (PN): $\Theta \sim PN(\boldsymbol{\mu}_{wi},\boldsymbol{\Sigma}_{wi})$.  
Let $\mathbf{U}_i = (U_{i1},U_{i2})^{\prime}$, where $U_{i1} = \cos \Theta_i$ and $U_{i2} = \sin \Theta_i$, the following explicit relation exists between $\mathbf{W}_i$ and $\Theta_i$: 
\begin{equation} 	\label{eq:u}
	\mathbf{W}_i=R_{i} \left(
	\begin{array}{c}
		\cos \Theta_i\\
		\sin \Theta_i
	\end{array}
	\right) = R_{i}\mathbf{U}_i, \, R_i= ||\mathbf{W}_i ||.
\end{equation}
The couple $(\Theta_i, R_i)$ is the representation  in polar coordinates of $\mathbf{W}_i$.

A natural way to define an $n-$variate  projected normal is to consider a $2n$-dimensional vector   $\mathbf{W}= \{ \mathbf{W}_{i} \}_{i=1}^{n}$ distributed as a $2n$-variate normal with mean vector   $\boldsymbol{\mu}_w$ and covariance matrix $\boldsymbol{\Sigma}_w$. The random  vector $\boldsymbol{\Theta}= \{\Theta_i\}_{i=1}^n$, of associated circular variables, is said to be distributed as an $n-$variate projected normal ($PN_n$): $\boldsymbol{\Theta} \sim PN_{n}(\boldsymbol{\mu}_w, \boldsymbol{\Sigma}_w)$.

The projected normal distribution is often considered in a univariate setting. Multivariate extensions have been developed in a spatial or spatio-temporal framework only, see for example \cite{wang2014} or \cite{mastrantonio2015b}.

\subsection{The skew normal} \label{sec:skew}


To model the linear part of telemetry data, we consider a skew normal distribution. 
Let $\boldsymbol{Y} = \{Y_{j} \}_{j=1}^q$ be a $q-$variate random variable, let $\boldsymbol{\mu}_{y}$ be a vector of length $q$, $\boldsymbol{\Sigma}_y$ be a $q \times q$ covariance matrix and $\boldsymbol{\Lambda}$ be a $q \times q$ matrix with elements belonging to $\mathbb{R}$. $\boldsymbol{Y}$ is distributed as a $q-$variate skew normal \citep{sahu2003} with parameters $\boldsymbol{\mu}_{y}$,  $\boldsymbol{\Sigma}_y$ and $\boldsymbol{\Lambda}$ ($\boldsymbol{Y} \sim SN_{q}(\boldsymbol{\mu}_{y}, \boldsymbol{\Sigma}_y,\boldsymbol{\Lambda})$) and   it has probability density function (pdf) 
\begin{equation} \label{eq:skew}
	2^q\phi_q \left(\mathbf{y}| \boldsymbol{\mu}_{y}, \boldsymbol{\Upsilon} \right) \Phi_q \left(\boldsymbol{\Lambda}^{\prime}\boldsymbol{\Upsilon}^{-1}(\mathbf{y}-\boldsymbol{\mu}_y)|\mathbf{0}_q, \boldsymbol{\varGamma}  \right),
\end{equation}
where $\phi_q(\cdot| \cdot, \cdot)$  and $\Phi_q(\cdot| \cdot, \cdot)$ indicate respectively the $q-$variate normal pdf and   cumulative distribution function, $\mathbf{0}_q$ is a vector of 0s of dimension $q$, $\boldsymbol{\Upsilon} = \boldsymbol{\Sigma}_{y}+\boldsymbol{\Lambda}\boldsymbol{\Lambda}^{\prime}$ and $\boldsymbol{\varGamma} = \mathbf{I}_q-\boldsymbol{\Lambda}^{\prime}\boldsymbol{\Upsilon}^{-1}\boldsymbol{\Lambda}$. The parameter $\boldsymbol{\Lambda}$ is generally called the \emph{skew parameter} and if all its elements  are 0,  then $\mathbf{Y} \sim N_q \left( \boldsymbol{\mu}_{y},\boldsymbol{\Sigma}_{y} \right)$.

The skew normal distribution has a nice stochastic representation, that follows from Proposition 1 of \cite{arellano2007}.  Let  $\mathbf{D} \sim HN_q(\mathbf{0}, \mathbf{I}_q)$, where $HN_q(\cdot,\cdot)$ indicates the $q-$dimensional half normal distribution, and  $\mathbf{H} \sim N_q(\mathbf{0}, \boldsymbol{\Sigma}_{y})$, then  
\begin{equation} \label{eq:skewrep}
	\mathbf{Y} = \boldsymbol{\mu}_{y} + \boldsymbol{\Lambda} \mathbf{D}+\mathbf{H},
\end{equation}
and $Y \sim SN_{q}(\boldsymbol{\mu}_{y}, \boldsymbol{\Sigma}_y,\boldsymbol{\Lambda})$. 
The mean vector and covariance matrix of $\mathbf{Y} $ are given by:
\begin{align} 
	E(\mathbf{Y}) & =\boldsymbol{\mu}_{y}+ \boldsymbol{\Lambda}\sqrt{\frac{2}{\pi}}, \label{eq:mean} \\
	\mbox{Var}(\mathbf{Y}) & =\boldsymbol{\Sigma}_{y}+ \left(1+ \frac{2}{\pi}\right)  \boldsymbol{\Lambda}\boldsymbol{\Lambda}^{\prime}. \label{eq:var}
\end{align}
In the general case, the  (multivariate or univariate) marginal distributions   of  $\mathbf{Y}$ are not skew normal \citep{sahu2003} but if  $\boldsymbol{\Lambda} = \mbox{diag}(\boldsymbol{\lambda})$, where $\boldsymbol{\lambda}=\{\lambda_{i}  \}_{i=1}^q$,  then all the marginal distributions  are  skew normal and  $\lambda_{i}$  affects only the  mean and variance of $Y_{i}$.

\subsection{The joint linear-circular distribution} \label{sec:joint}

The new multivariate circular-linear distribution  proposed is obtained as follows.  Let  $$(\mathbf{W},\mathbf{Y})^{\prime}\sim SN_{2n+q}(\boldsymbol{\mu}, \boldsymbol{\Sigma},\text{diag}((\mathbf{0}_{2n},\boldsymbol{\lambda}))),$$
with  $\boldsymbol{\mu} = (\boldsymbol{\mu}_w,\boldsymbol{\mu}_y)$ and $ 
\boldsymbol{\Sigma}=\left(
\begin{array}{cc}
\boldsymbol{\Sigma}_{w} & \boldsymbol{\Sigma}_{wy}\\
\boldsymbol{\Sigma}_{wy}^{\prime} & \boldsymbol{\Sigma}_{y}
\end{array}
\right),
$ 
where $\boldsymbol{\Sigma}$ is a $(2n+q)\times (2n+q) $  covariance matrix.
The marginal distribution of $\mathbf{W}$ is  a $2n$-variate normal  with mean $\boldsymbol{\mu}_w$ and covariance matrix $\boldsymbol{\Sigma}_w$, since the associate skew parameters are all zeros, while $\mathbf{Y}\sim SN_{q}(\boldsymbol{\mu}_y, \boldsymbol{\Sigma}_y, \text{diag}(\boldsymbol{\lambda}))$.

If we  apply the transformation \eqref{eq:tranpn} to the components $\mathbf{W}_i$ of $(\mathbf{W},\mathbf{Y})$, then $(\boldsymbol{\Theta}, \mathbf{Y})^{\prime}$ 
is a multivariate vector of $n$ circular and $q$ linear variables and we say that is distributed as an $(n,q)$-variate joint projected and skew normal ($JPSN_{n,q}$) with parameters $\boldsymbol{\mu}$, $\boldsymbol{\Sigma}$ and $\boldsymbol{\lambda}$: $(\boldsymbol{\Theta},\mathbf{Y})^{\prime}\sim JPSN_{n,q}(\boldsymbol{\mu}, \boldsymbol{\Sigma}, \boldsymbol{\lambda})$. A closed form for the joint distribution is available by introducing suitable latent variables (see Section \ref{sec:idsolve})

As the skew matrix is  diagonal, each marginal distribution of $(\mathbf{W},\mathbf{Y})^{\prime}$ is still a skew normal  with parameters given by the appropriate subset  of $\boldsymbol{\mu}$, $\boldsymbol{\Sigma}$ and $\boldsymbol{\lambda}$. Accordingly  all the marginal distributions of $(\boldsymbol{\Theta},\mathbf{Y})^{\prime}$  are joint projected and skew normals.

The interpretation of the parameters $\boldsymbol{\mu}_{y}$,  $\boldsymbol{\Sigma}_y$ and $\boldsymbol{\lambda}$ is straightforward. The interpretation of  $(\boldsymbol{\mu}_{wi}, \boldsymbol{\Sigma}_{wi})^{\prime}$, i.e. the parameters of the univariate marginal projected distribution, is not easy because there is a complex interaction between them  and it is not clear how a single component of $\boldsymbol{\mu}_w$ or $\boldsymbol{\Sigma}_w$ affects the shape of the univariate density, that can be symmetric, asymmetric, unimodal or bimodal (for a discussion see \cite{Wang2013}). 
However, in a Bayesian framework we can compute Monte Carlo approximations of all the features of the marginal univariate circular distribution \citep{mastrantonio2015}, such as the directional mean, the circular concentration and the posterior predictive density, bypassing the difficult in the interpretation of the parameters  $(\boldsymbol{\mu}_{wi}, \boldsymbol{\Sigma}_{wi})^{\prime}$.

The two components of $\mathbf{U}_i$ are respectively the cosine and sine of the circular variable $\Theta_i$, see \eqref{eq:u}, and 
the correlation matrix of $(\mathbf{W}, \mathbf{Y})^{\prime}$, $\boldsymbol{\Omega}$, 
is the same of $(\mathbf{U}, \mathbf{Y})^{\prime}$, where $\mathbf{U}= \{ \mathbf{U}_i \}_{i=1}^n$. We can easily interpret the circular-circular and circular-linear dependence in terms of the correlation between the linear variables and the  sine and cosine of the circular ones.

The parameters of the  JPSN  are not identifiable, since   $\mathbf{W}_i$ and   $c_i \mathbf{W}_i$, with $c_i>0$, produce the same $\Theta_i$, and hence the same $\boldsymbol{\Theta}$. As  consequence the distribution of $(\boldsymbol{\Theta},\mathbf{Y})^{\prime}$ is unchanged if the parameters $(\boldsymbol{\Sigma}, \boldsymbol{\mu},\boldsymbol{\lambda})$ are replaced  by $\left(\mathbf{C}\boldsymbol{\mu},\mathbf{C} \boldsymbol{\Sigma}\mathbf{C},\boldsymbol{\lambda}\right)$, where $\mathbf{C} = \mbox{diag}(\mathbf{c},\mathbf{1}_q) $ with  $\mathbf{c}= \{ (c_i,c_i)^{\prime} \}_{i=1}^n$; to identify the parameters a constraint is needed. Without loss of generality, following and extending  \cite{Wang2013}, we can fix the scale of each $\mathbf{W}_i$ by setting to a constant, say 1, each second element of the diagonals of the $\boldsymbol{\Sigma}_{wi}$s. The constrains create some  difficult in the estimation of  $\boldsymbol{\Sigma}$ since we have to ensure that it is a positive definite (PD) matrix.
To avoid confusion, from now to go on we indicate with $\tilde{\boldsymbol{\Sigma}}$ and $\tilde{\boldsymbol{\mu}}$ the identifiable version of $\boldsymbol{\Sigma}= \mathbf{C}\tilde{\boldsymbol{\Sigma}}\mathbf{C}$ and $\boldsymbol{\mu}= \mathbf{C}\tilde{\boldsymbol{\mu}}$.

\section{The Bayesian inference} \label{sec:idsolve}

Suppose to have $T$ observations drawn from an $(n,q)$-variate JPSN, $(\boldsymbol{\Theta}_t,\mathbf{Y}_t)^{\prime} \sim JPSN_{n,q}(\tilde{\boldsymbol{\mu}},\tilde{\boldsymbol{\Sigma}},\boldsymbol{\lambda} )$ with $t=1,\dots , T$, where $\boldsymbol{\Theta}_t=\{ \Theta_{ti} \}_{i=1}^n$ and $\mathbf{Y}_t = \{ Y_{tj}\}_{j=1}^q$. With a slight abuse of notation, let $\boldsymbol{\Theta} = \{  \boldsymbol{\Theta}_t \}_{t=1}^T$ and $\mathbf{Y}= \{ \mathbf{Y}_t \}_{t=1}^T$ and suppose that 
%
%
given  $({\boldsymbol{\Theta}},{\mathbf{Y}})$ we want to learn about $\tilde{\boldsymbol{\mu}}$, $\tilde{\boldsymbol{\Sigma}}$ and $\boldsymbol{\lambda}$ in a Bayesian perspective using an MCMC algorithm, i.e. obtaining  samples from the posterior distribution  
\begin{equation} \label{eq:post}
	f(\tilde{\boldsymbol{\mu}}, \tilde{\boldsymbol{\Sigma}}, \boldsymbol{\lambda}| \boldsymbol{\theta},\mathbf{y}) \propto \prod_{t=1}^T f( \boldsymbol{\theta}_t,\mathbf{y}_t|\tilde{\boldsymbol{\mu}}, \tilde{\boldsymbol{\Sigma}},\boldsymbol{\lambda})f(\tilde{\boldsymbol{\mu}}, \tilde{\boldsymbol{\Sigma}},\boldsymbol{\lambda}).
\end{equation}

We cannot work directly with the posterior \eqref{eq:post} since $f( \boldsymbol{\theta}_t,\mathbf{y}_t|\tilde{\boldsymbol{\mu}}, \tilde{\boldsymbol{\Sigma}},\boldsymbol{\lambda})$ is not known in closed form  and it is not easy to find an appropriate  prior distribution $f(\tilde{\boldsymbol{\mu}}, \tilde{\boldsymbol{\Sigma}},\boldsymbol{\lambda})$, even if we assume independence between the parameters, i.e. $f(\tilde{\boldsymbol{\mu}}, \tilde{\boldsymbol{\Sigma}},\boldsymbol{\lambda}) = f(\tilde{\boldsymbol{\mu}})f( \tilde{\boldsymbol{\Sigma}})f(\boldsymbol{\lambda})$, because  $f( \tilde{\boldsymbol{\Sigma}})$ must be  a valid distribution for a PD matrix with some of its diagonal elements constrained. 
%
%
%
%

To solve both problems let $\mathbf{W}_{ti}$ be the bivariate linear variable associated with $\Theta_{ti}$, let $R_{ti}= ||\mathbf{W}_{ti}||$ and let $\mathbf{D}_t$ be the  $q-$variate half normal random variable associated with $\mathbf{Y}_t$ in the stochastic representation given in     \eqref{eq:skewrep}.  Let $\mathbf{R}_t= \{R_{ti}\}_{i=1}^n$, $\mathbf{R}=\{\mathbf{R}_t\}_{t=1}^T$ and, again with a slight abuse of notation,  $\mathbf{D}= \{ \mathbf{D}_t \}_{t=1}^T$. Instead of the posterior \eqref{eq:post} we evaluate, i.e. we obtain posterior samples, from the posterior
\begin{equation}  \label{eq:2}
	f({\boldsymbol{\mu}}, {\boldsymbol{\Sigma}},\boldsymbol{\lambda}, {\mathbf{r}},{\mathbf{d}}| \boldsymbol{\theta},\mathbf{y}) \propto \prod_{t=1}^T f(\boldsymbol{\theta}_t,{\mathbf{r}}_t,\mathbf{y}_t,{\mathbf{d}}_t|{\boldsymbol{\mu}}, {\boldsymbol{\Sigma}},\boldsymbol{\lambda}) f({\boldsymbol{\mu}}, {\boldsymbol{\Sigma}},\boldsymbol{\lambda}),
\end{equation}
where the joint density of $(\boldsymbol{\Theta}_t, \mathbf{R}_t, \mathbf{Y}_t,\mathbf{D}_t)$ is 
\begin{equation} \label{eq:tot1}
	2^q\phi_{2n+q}((\mathbf{w}_t, \mathbf{y}_t-\text{diag}(\boldsymbol{\lambda})\mathbf{d}_t )^{\prime} |{\boldsymbol{\mu}}, {\boldsymbol{\Sigma}}) \phi_q(\mathbf{d}_t| \mathbf{0},\mathbf{I}_q ) \prod_{i=1}^n r_{ti}.
\end{equation}
Equation   \eqref{eq:tot1} is the density that arises by transforming each $\mathbf{W}_{ti}$, in the joint density of $(\mathbf{W}_t, \mathbf{Y}_t,\mathbf{D}_t)$, to its representation in polar coordinate (equation \eqref{eq:u}). 
Note that in  \eqref{eq:2} we work with the unconstrained PD matrix ${\boldsymbol{\Sigma}}
$ and then the definition of the prior distribution is easier with respect to \eqref{eq:post}.  The posterior distribution \eqref{eq:2}  is not identifiable, but nevertheless we can obtain samples from it. Suppose to have   $B$ samples from \eqref{eq:2}, i.e.  
$\{{\boldsymbol{\mu}}^b, {\boldsymbol{\Sigma}}^b,\boldsymbol{\lambda}^b,{\mathbf{r}}^b,{\mathbf{d}}^b\}_{b=1}^B$. The subset $\{{\boldsymbol{\mu}}^b, {\boldsymbol{\Sigma}}^b,\boldsymbol{\lambda}^b\}_{b=1}^B$ are samples from the posterior distribution $f({\boldsymbol{\mu}}, {\boldsymbol{\Sigma}},\boldsymbol{\lambda}| \boldsymbol{\theta},\mathbf{y})$ and
if we transform the set $\{{\boldsymbol{\mu}}^b, {\boldsymbol{\Sigma}}^b,\boldsymbol{\lambda}^b\}_{b=1}^B$ to the set $\{\tilde{\boldsymbol{\mu}}^b, \tilde{\boldsymbol{\Sigma}}^b,\mathbf{C}^b,\boldsymbol{\lambda}^b\}_{b=1}^B$,  the latter is a set of samples from $f(\tilde{\boldsymbol{\mu}}, \tilde{\boldsymbol{\Sigma}}, \mathbf{C},\boldsymbol{\lambda}| \mathbf{u},\mathbf{y})$. As consequence the subset $\{\tilde{\boldsymbol{\mu}}^b, \tilde{\boldsymbol{\Sigma}}^b,\boldsymbol{\lambda}^b\}_{b=1}^B$ are samples from \eqref{eq:post}, the posterior distribution of interest.  From a practical point  of view,  we can work with \eqref{eq:2} and  put a prior distribution over $({\boldsymbol{\mu}}, {\boldsymbol{\Sigma}}, \boldsymbol{\lambda})$. The posterior samples $\{{\boldsymbol{\mu}}^b, {\boldsymbol{\Sigma}}^b,\boldsymbol{\lambda}^b\}_{b=1}^B$ are transformed to the set $\{\tilde{\boldsymbol{\mu}}^b, \tilde{\boldsymbol{\Sigma}}^b,\boldsymbol{\lambda}^b\}_{b=1}^B$ that are  posterior samples from  \eqref{eq:post}. 
The prior distribution $f(\tilde{\boldsymbol{\mu}}, \tilde{\boldsymbol{\Sigma}},\boldsymbol{\lambda})$ in \eqref{eq:post} is induced by $f({\boldsymbol{\mu}}, {\boldsymbol{\Sigma}},\boldsymbol{\lambda})$ in  \eqref{eq:2}. 
To verify what is the real advantage of this approach, let assume  $f({\boldsymbol{\mu}}, {\boldsymbol{\Sigma}}, \boldsymbol{\lambda}) = f({\boldsymbol{\mu}}, {\boldsymbol{\Sigma}})f( \boldsymbol{\lambda})$. The full conditional  of  $({\boldsymbol{\mu}}, {\boldsymbol{\Sigma}}) $ is  proportional to 
$ 
\prod_{t=1}^T\phi_{2n+q}((\mathbf{w}_t, \mathbf{y}_t -\text{diag}(\boldsymbol{\lambda}) \mathbf{d}_t )^{\prime}|\boldsymbol{\mu}, \boldsymbol{\Sigma}) f({\boldsymbol{\mu}}, {\boldsymbol{\Sigma}}), 
$ 
i.e. the product of  a $(2n+q)$-variate normal density and a prior distribution over its mean and covariance matrix. We can then use the standard prior for the normal likelihood that gives the possibility to find in closed form the full conditional of $(\boldsymbol{\mu}, \boldsymbol{\Sigma})$.

We suggest
${\boldsymbol{\mu}},{\boldsymbol{\Sigma}} \sim NIW(\cdot,\cdot,\cdot, \cdot )$, where $NIW(\cdot,\cdot,\cdot, \cdot )$ indicates the normal inverse Wishart (NIW) distribution. This induces a   full conditional for $({\boldsymbol{\mu}}, {\boldsymbol{\Sigma}})$ that is NIW and then it is easy to simulate with a Gibbs step. We can apply our approach to obtain posterior samples with a Gibbs step even when we have only circular variables, i.e. we are working with the multivariate projected normal, and also  in the univariate case where,  till now, the components of $\boldsymbol{\Sigma}_{wi}$ were sampled   using  Metropolis steps, see for example \cite{Wang2013}, \cite{wang2014}, \cite{mastrantonio2015} or \cite{mastrantonio2015b}. Under the NIW we are not able to compute, in closed form, the induced prior on $(\tilde{\boldsymbol{\mu}},\tilde{\boldsymbol{\Sigma}})$ but, if  needed, it can always be evaluated through simulation. Of course the NIW it is not the only possible choice,  for example can be used the prior proposed by 
\cite{Huang2013} or the one of \cite{omalley2008}, but we think that the NIW  is easiest to implement.

%
%
%
%

To conclude the  MCMC specification, we have to show how to sample the remaining parameters and latent variables.
Let $\boldsymbol{\mu}_{y_t|w_t} =\boldsymbol{\mu}_{y}+\boldsymbol{\Sigma}_{wy}^{\prime}\boldsymbol{\Sigma}_{w}^{-1}\left( \mathbf{w}_t - \boldsymbol{\mu}_{w} \right) $ and   $\boldsymbol{\Sigma}_{y|w} = \boldsymbol{\Sigma}_{y}- \boldsymbol{\Sigma}_{wy}^{\prime}\boldsymbol{\Sigma}_{w}^{-1}\boldsymbol{\Sigma}_{wy}$. The full conditional of $\boldsymbol{\lambda}$ is proportional to 
$\prod_{t=1}^T \phi_{q}(\mathbf{y}_t |\boldsymbol{\mu}_{y_t|w_t}+ \text{diag}(\mathbf{d}_t) \boldsymbol{\lambda}   , \boldsymbol{\Sigma}_{y|w} ) f(\boldsymbol{\lambda})$
and if we use a multivariate normal prior over $\boldsymbol{\lambda}$, we obtain a multivariate normal full conditional. Let $\mathbf{V}_{d} =\left(\boldsymbol{\Lambda}^{\prime} \boldsymbol{\Sigma}_{y|w}^{-1}\boldsymbol{\Lambda}+ \mathbf{I}_{q}  \right)^{-1} $ and $\mathbf{M}_{d_t} = \mathbf{V}_d\boldsymbol{\Lambda}^{\prime} \boldsymbol{\Sigma}_{y|w}^{-1}   \left(\mathbf{y}_{t}- \boldsymbol{\mu}_{y|w}\right)$ then  the full condition of $\mathbf{d}_t$ is $N_{q}\left(\mathbf{M}_{d_t}, \mathbf{V}_q  \right)I_{\mathbf{0_q, \boldsymbol{\infty}}}$, where $N_{q}\left(\cdot, \cdot  \right)I_{\mathbf{0_q, \boldsymbol{\infty}}}$ is a  $q-$dimensional truncated normal distribution with  components having support $\mathbb{R}^+$.
We are not able to find in closed form the full conditionals of the $r_{ti}$s  and then we  sample them with  Metropolis steps.

\section{The hidden Markov model} \label{sec:HPD-HMM}

In this Section,  we  introduce the HMM and  its Bayesian non-parametric version,  the sHDP-HMM.

\subsection{The model} \label{sec:HMM}
%

Let $z_t \in \mathcal{K}  \subseteq  \mathbb{Z}^+ \backslash \{ 0 \}$ be a discrete variable which indicates the latent behaviour  
at time $t$ and let 
$\boldsymbol{\psi}_{k}$ be the vector of  parameters of the JPSN  in the behaviour $k$, i.e.  $\boldsymbol{\psi}_{k}= (\boldsymbol{\mu}_k, \boldsymbol{\Sigma}_k, \boldsymbol{\lambda}_k)$.


In the HMM  the observations are time independent  given 
$\{z_t\}_{t=1}^T$ and $\{\boldsymbol{\psi}_{k}\}_{k \in \mathcal{K}}$, i.e.:
\begin{align}
	f(\boldsymbol{\theta}, \mathbf{y}|\{z_t\}_{t \in \mathcal{T}} , \{\boldsymbol{\psi}_{k}\}_{k \in \mathcal{K}}) & =   \prod_{t \in \mathcal{T}} \prod_{k \in \mathcal{K}} \left[  f(\boldsymbol{\theta}_t, \mathbf{y}_t|\boldsymbol{\psi}_{z_t})  \right]^{I(z_t,k)} ,
\end{align}  
where $I(z_t,k)$,  the indicator function, is equal to 1  if $z_t=k$, 0 otherwise. The hidden variables $\{z_{t}\}_{t=1}^T$  follow a first-order Markov chain with $P(z_t=k|z_{t-1}=j)= \pi_{jk}$ and  
$$
z_t | z_{t-1} , \boldsymbol{\pi}_{z_{t-1}} \sim \boldsymbol{\pi}_{z_{t-1}},
$$
where $\boldsymbol{\pi}_{j} = \{ {\pi}_{jk}\}_{k \in \mathcal{K}}$.  As pointed out by \cite{cappe2005}, the initial state $z_0$ cannot be estimated consistently since we have no observation  at time 0 and then we set $z_0=1$. 
The distribution of  $\boldsymbol{\Theta}_t, \mathbf{Y}_t|\boldsymbol{\psi}_{z_t}$, that in the HMM literature it is called the \emph{emission distribution}, is the JPSN, i.e.  $\boldsymbol{\Theta}_t, \mathbf{Y}_t|z_t, \boldsymbol{\psi}_{z_t} \sim JPSN_{n,q}(\boldsymbol{\mu}_{z_t}, \boldsymbol{\Sigma}_{z_t}, \boldsymbol{\lambda}_{z_t})$.
%
%
%
We remark that, although the model is specified with respect to $(\boldsymbol{\mu}_{k}, \boldsymbol{\Sigma}_{k})$, we can only estimate $(\tilde{\boldsymbol{\mu}}_{k}, \tilde{\boldsymbol{\Sigma}}_{k})$


We can equivalently express  the hidden process in a    more suitable  way for the specification of the sHDP-HMM.  Let $\boldsymbol{\eta}_t = \boldsymbol{\psi}_{z_t}$, with $ \boldsymbol{\psi}_k \in \boldsymbol{\Psi} ,\,k \in \mathcal{K}$,  and suppose that each element of the sequence $\{  \boldsymbol{\eta}_t\}_{t \in \mathcal{T}}$,  is drawn from a  discrete space $\boldsymbol{\Xi}=\{ \boldsymbol{\psi}_k \}_{k \in \mathcal{K}}$ and  the probability of drawing  $  \boldsymbol{\eta}_t$ depends only on the value $\boldsymbol{\eta}_{t-1}$.  We let  $P(\boldsymbol{\eta}_{t}|\boldsymbol{\eta}_{t-1}=\boldsymbol{\psi}_{l}) \sim G_{\boldsymbol{\eta}_{t-1} } \equiv  G_{\boldsymbol{\psi}_{l} }$, with $G_{\boldsymbol{\psi}_{l} } = \sum_{k \in \mathcal{K}} \pi_{lk}\delta_{\boldsymbol{\psi}_k}$  
where $\delta_{\boldsymbol{\psi}_k}$ is the \emph{Dirac delta function} placed on $\boldsymbol{\psi}_k$. The above HMM can be expressed as 
\begin{align}
	f(\boldsymbol{\theta}, \mathbf{y}|\{\boldsymbol{\eta}_t\}_{t \in \mathcal{T}}) & =   \prod_{t \in \mathcal{T}} f(\boldsymbol{\theta}_t, \mathbf{y}_t|\boldsymbol{\eta}_t),  \\ 
	\boldsymbol{\Theta}_t, \mathbf{Y}_t|\boldsymbol{\eta}_t &\sim JPSN_{n,q}(\boldsymbol{\mu}_{z_t}, \boldsymbol{\Sigma}_{z_t}, \boldsymbol{\lambda}_{z_t}),\\
	\boldsymbol{\eta}_t | \boldsymbol{\eta}_{t-1},G_{\eta_{t-1}} &\sim G_{\eta_{t-1}}.
\end{align}

The standard way to estimate the  cardinality of $\mathcal{K}$ $(K^*)$
is to set it a priori  and then run models with different values of $K^*$.
The models are compared  using informational criteria such as the AIC, BIC or ICL and  the model that has the better value of the selected  criterion is chosen. They can suggest different values of $K^*$ and moreover they are used to obtain the optimal $K^*$ but without  any measurement of uncertainty.

In a Bayesian framework there are several ways to deal with an unknown number of behaviours. We can  use the Reversible Jump proposed by \cite{GREEN1995} or the HDP by \cite{Teh2006} and its modification, the sticky hierarchical Dirichlet process (sHDP), proposed by \cite{fox2011}. Here we use the sHDP because it is easier to implement. This method let $K^* \rightarrow \infty$ and estimates from the data the  number of non-empty behaviours, $K$.  $K$ is a random variable and   we can have a measurement of uncertainty on its estimate.


\subsection{The sHDP-HMM} \label{sec:stick1}

In the sHDP-HMM  is assumed the following:
\begin{align} \label{eq:G_0}
	G_{\boldsymbol{\eta}_t} | \tau,\rho, \gamma, H &\sim sHDP( \tau, \gamma, \rho, H) , \quad \tau>0,\gamma>0, \rho \in [0,1] ,  
\end{align}
where with $sHDP(\cdot)$ we indicate the sticky hierarchical Dirichlet process \citep{fox2011} with first level concentration parameter $\tau$, second level concentration parameter $\gamma$, self-transition parameter $\rho$ and base measure $H$, where the base measure is  a distribution over the space $\boldsymbol{\Psi}$. 
\cite{fox2011} show that equation \eqref{eq:G_0} can be written equivalently as 
\begin{align}
	G_{\boldsymbol{\eta}_t} | \rho, \gamma &\sim DP(\gamma,(1-\rho) G_0+\rho\delta_{\boldsymbol{\eta}_t})  \label{eq:G_j},\\
	G_0 | \tau, H &\sim DP(\tau, H) ,
\end{align}
where  $DP(v,L)$ indicates the Dirichlet process with base measure $L$ and 
concentration parameter $a$.

We can write the sHDP-HMM  as
\begin{align}
	f(\boldsymbol{\theta}, \mathbf{y}|\{\boldsymbol{\eta}_t\}_{t \in \mathcal{T}}) & =   \prod_{t \in \mathcal{T}} f(\boldsymbol{\theta}_t, \mathbf{y}_t|\boldsymbol{\eta}_t), \label{eq:like} \\ 
	\boldsymbol{\Theta}_t, \mathbf{Y}_t|\boldsymbol{\eta}_t &\sim JPSN_{n,q}(\boldsymbol{\mu}_{z_t}, \boldsymbol{\Sigma}_{z_t}, \boldsymbol{\lambda}_{z_t}),\\
	\boldsymbol{\eta}_t | \boldsymbol{\eta}_{t-1},G_{\eta_{t-1}} &\sim G_{\eta_{t-1}},\\
	G_{\boldsymbol{\eta}_t} | \rho, \gamma &\sim DP(\gamma,(1-\rho) G_0+\rho\delta_{\boldsymbol{\eta}_t}) , \label{eq:G_j2}\\
	G_0 | \tau, H &\sim DP(\tau, H) .
\end{align}

To simplify the implementation we write the model using the stick-breaking representation of the Dirichlet process \citep{sethuraman:stick} and we introduce the latent variables $\mathbf{r}= \{ \mathbf{r}_t \}_{t=1}^T$ and $\mathbf{d}= \{  \mathbf{d}_t \}_{t=1}^T$. Then, 
let  $\boldsymbol{\beta}=\{\beta_k\}_{k=1}^{\infty}$,  the model we estimate with the MCMC algorithm is
\begin{align}
	f(\boldsymbol{\theta}, \mathbf{y},\mathbf{r},\mathbf{d}|\{z_t\}_{t \in \mathcal{T}} , \{\boldsymbol{\psi}_{k}\}_{k \in \mathcal{K}}) & =   \prod_{t \in \mathcal{T}} \prod_{k \in \mathcal{K}} \left[  f(\boldsymbol{\theta}_t,\mathbf{r}_t, \mathbf{y}_t,\mathbf{d}_t|\boldsymbol{\psi}_{k})  \right]^{I({k=z_t})}, \label{eq:stick2}\\
	z_t | z_{t-1} , \boldsymbol{\pi}_{z_{t-1}}& \sim \boldsymbol{\pi}_{z_{t-1}},\\
	\boldsymbol{\pi}_{{k}}  | \rho,\gamma,\boldsymbol{\beta},\boldsymbol{\psi}_{k}& \sim DP\left(\gamma, (1-\rho) \boldsymbol{\beta}+\rho\delta_{\boldsymbol{\psi}_{k}} \right),\\
	\boldsymbol{\beta}|\tau &\sim GEM(\tau) ,\\
	\boldsymbol{\psi}_{k}| H &\sim H ,  \label{eq:H}
\end{align}
where $GEM(\cdot)$ indicates the stick-breaking process \citep{sethuraman:stick}.

To complete the model we have to specify the base measure $H$, that in the stick-breaking representation acts as a prior  distribution over the parameters $(\boldsymbol{\mu}_{k}, \boldsymbol{\Sigma}_{k}, \boldsymbol{\lambda}_{k})$. We use $\boldsymbol{\mu}_{k}, \boldsymbol{\Sigma}_{k} \sim NIW({\boldsymbol{\mu}}_{0},{\eta},{\varsigma}, {\boldsymbol{\Psi}} )$ 
and  $\boldsymbol{\lambda} \sim N_q({\mathbf{M}},{\mathbf{V}})$ because, as noted in Section \ref{sec:joint},  these choices  lead to full conditionals  easy to simulate.  

We assume that the parameters of the sHDP,  $\tau$, $\gamma$ and $\rho$, are random quantities and following \cite{fox2011} we  choose as priors: $\tau \sim G(a_{\tau}, b_{\tau})$, $\gamma \sim G(a_{\gamma}, b_{\gamma})$ and $\rho \sim B(a_{\rho}, b_{\rho})$, where $G(\cdot, \cdot)$ indicates the gamma distribution, expressed in terms of  shape and scale, and $B(\cdot, \cdot)$ is the beta.  Since we treat $\rho$ as a random variable we can estimate through the data the strength of the self-transition.
For the MCMC sampling of the behaviour indicator variables we use the \emph{beam sampler}  \citep{VanGael2008}. Despite the complexity  of the model, in terms of emission distribution and  the underline Markov structure,  with the exception of the $r_{it}$s, all the other unknown quantities can be updated in the MCMC with  Gibbs steps.



\section{Real data example} \label{sec:real}

\begin{figure}[t!]
	\centering
	{\subfloat[First dog]{\includegraphics[scale=0.24]{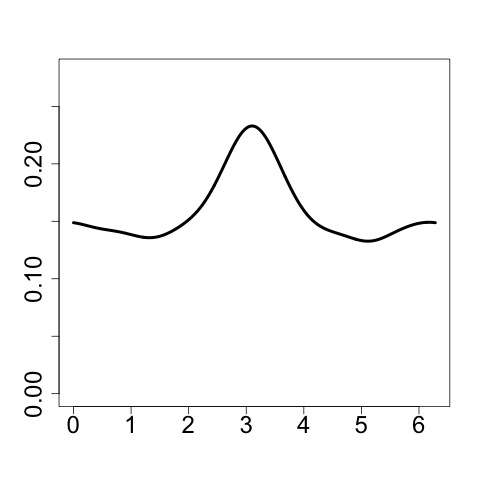}}}
	{\subfloat[Second dog]{\includegraphics[scale=0.24]{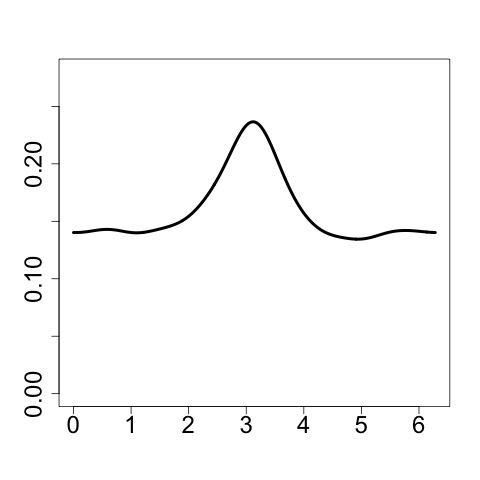}}}
	{\subfloat[Third dog]{\includegraphics[scale=0.24]{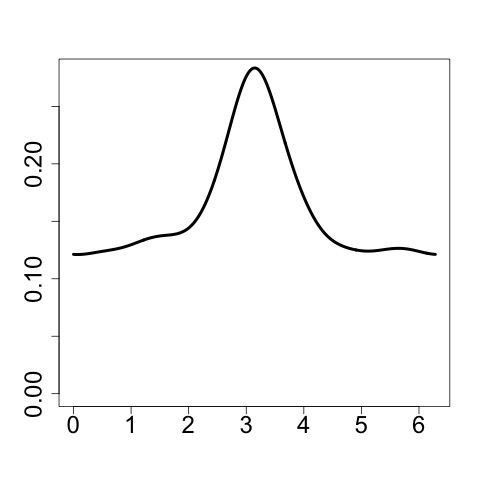}}}\\
	{\subfloat[Fourth dog]{\includegraphics[scale=0.24]{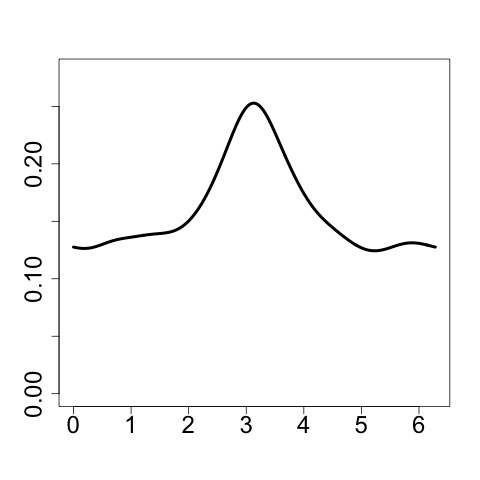}}}
	{\subfloat[Fifth dog]{\includegraphics[scale=0.24]{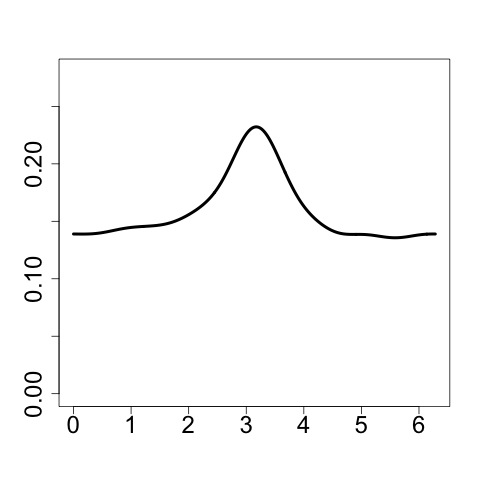}}}
	{\subfloat[Sixth dog]{\includegraphics[scale=0.24]{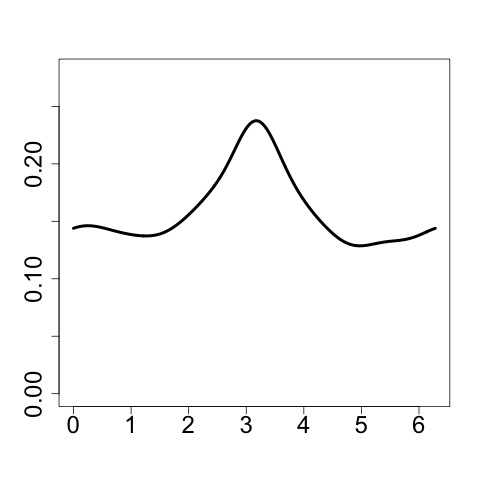}}}
	\caption{Marginal densities of the turning-angles.} \label{fig:Dcirc}
\end{figure}

\begin{figure}[t!]
	\centering
	{\subfloat[First dog]{\includegraphics[scale=0.24]{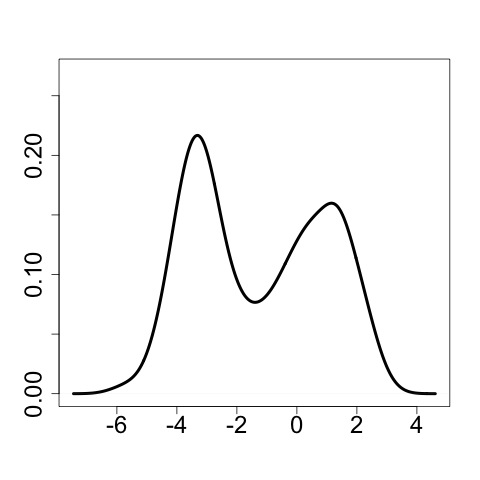}}}
	{\subfloat[Second dog]{\includegraphics[scale=0.24]{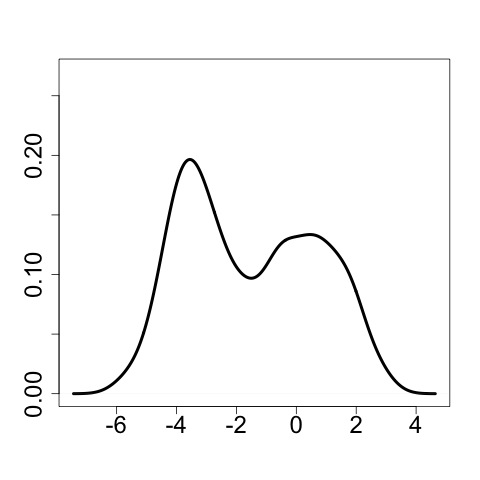}}}
	{\subfloat[Third dog]{\includegraphics[scale=0.24]{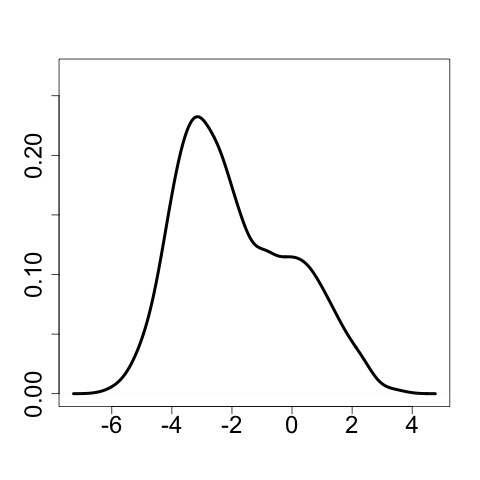}}}\\
	{\subfloat[Fourth dog]{\includegraphics[scale=0.24]{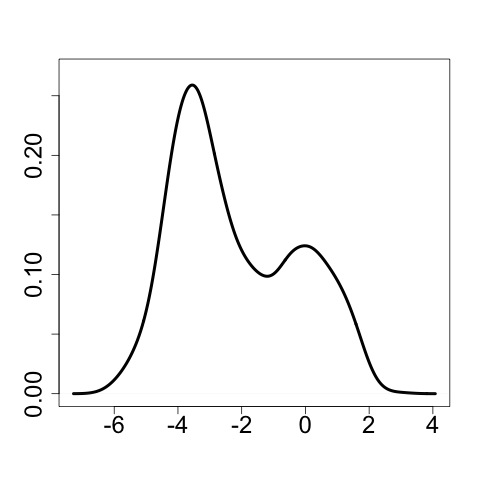}}}
	{\subfloat[Fifth dog]{\includegraphics[scale=0.24]{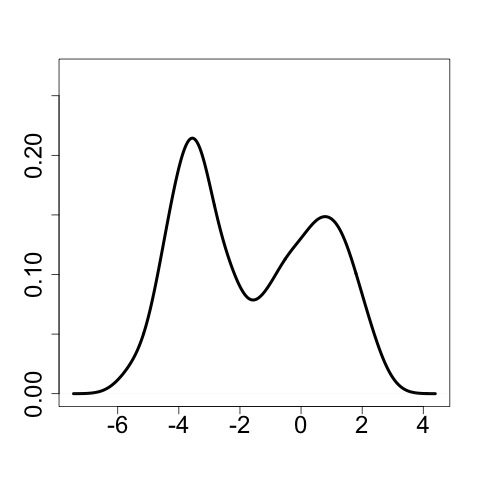}}}
	{\subfloat[Sixth dog]{\includegraphics[scale=0.24]{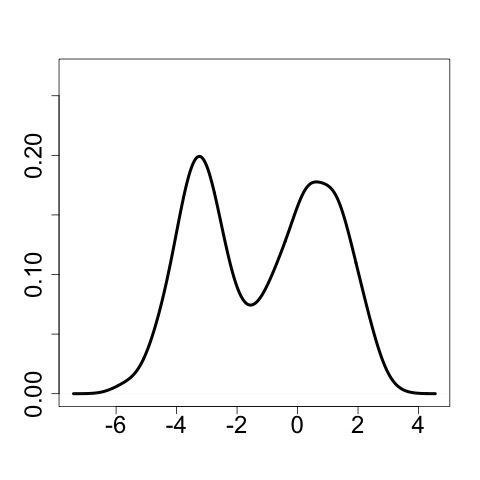}}}
	\caption{Marginal densities of the log-step-lengths.} \label{fig:Dlin}
\end{figure}

\begin{figure}[t!]
	\centering
	{\subfloat[First dog]{\includegraphics[scale=0.24]{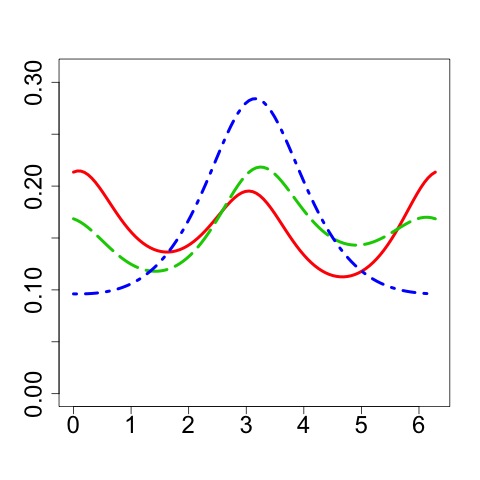}}}
	{\subfloat[Second dog]{\includegraphics[scale=0.24]{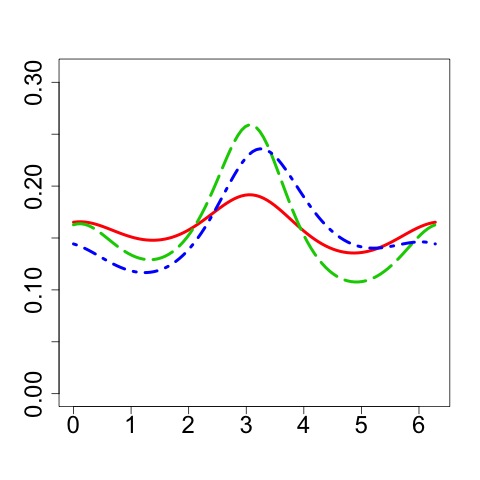}}}
	{\subfloat[Third dog]{\includegraphics[scale=0.24]{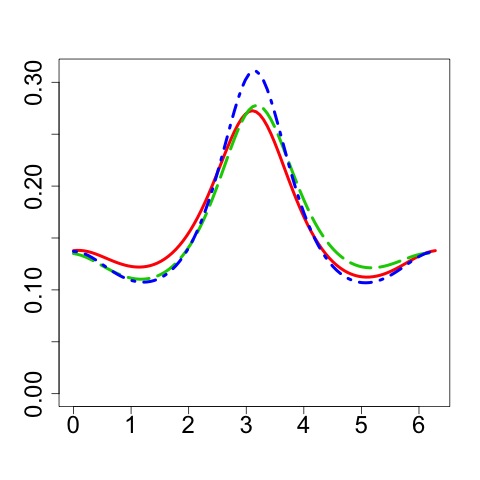}}}\\
	{\subfloat[Fourth dog]{\includegraphics[scale=0.24]{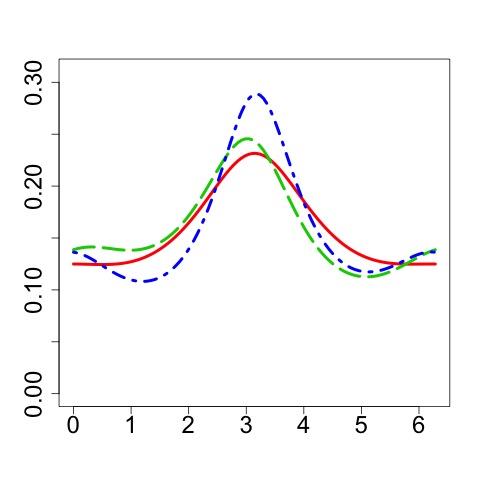}}}
	{\subfloat[Fifth dog]{\includegraphics[scale=0.24]{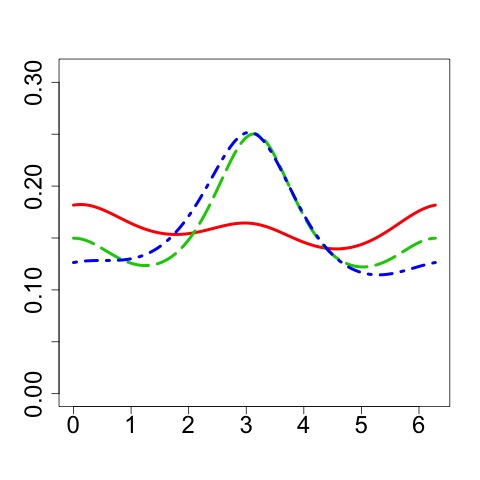}}}
	{\subfloat[Sixth dog]{\includegraphics[scale=0.24]{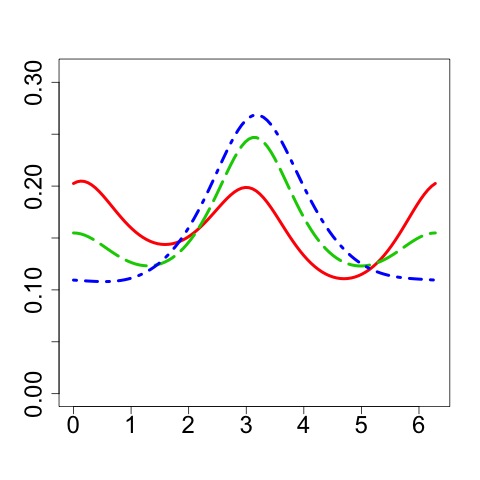}}}
	\caption{Posterior marginal densities of the turning-angles. The  dashed-dotted line  is  the marginal density of the  first behaviour, the dashed one is the second and the full the third. $\hat{K}=3$.
	} \label{fig:DestDir}
\end{figure}

\begin{figure}[t!]
	\centering
	{\subfloat[First dog]{\includegraphics[scale=0.24]{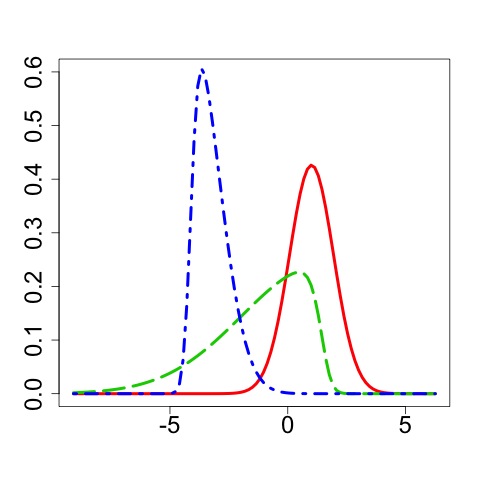}}}
	{\subfloat[Second dog]{\includegraphics[scale=0.24]{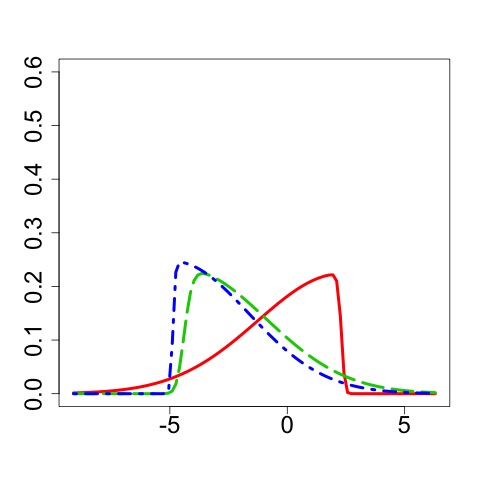}}}
	{\subfloat[Third dog]{\includegraphics[scale=0.24]{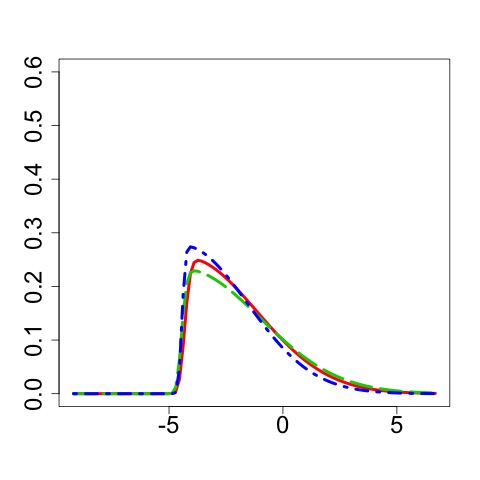}}}\\
	{\subfloat[Fourth dog]{\includegraphics[scale=0.24]{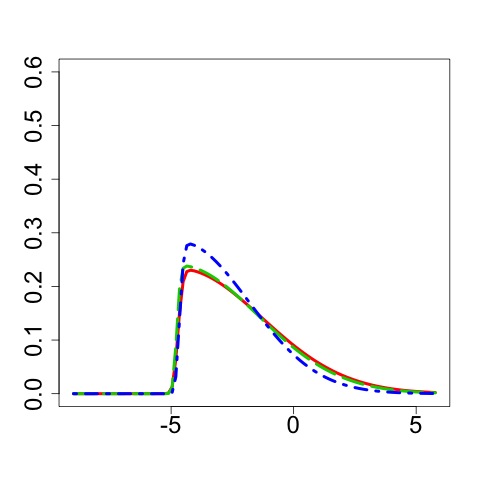}}}
	{\subfloat[Fifth dog]{\includegraphics[scale=0.24]{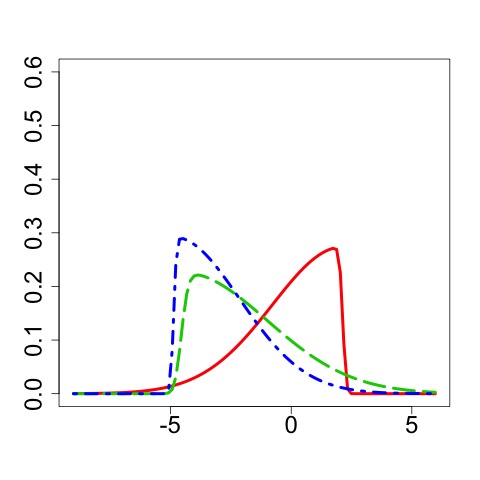}}}
	{\subfloat[Sixth dog]{\includegraphics[scale=0.24]{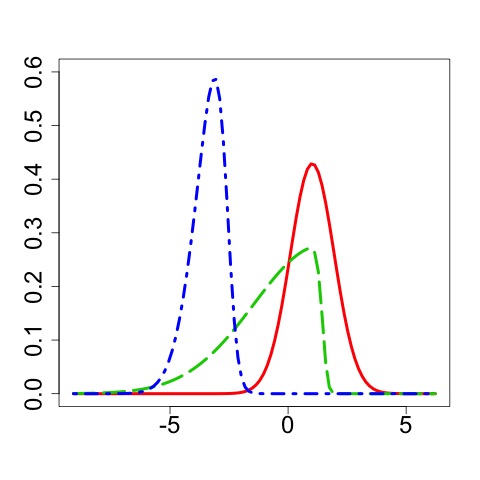}}}
	\caption{Posterior marginal densities of the log-step-lengths. The  dashed-dotted line  is  the marginal density of the  first behaviour, the dashed one is the second and the full the third. $\hat{K}=3$.
	} \label{fig:DestLin}
\end{figure}

\begin{figure}[t!]
	\centering
	{\subfloat[First behaviour]{\includegraphics[trim=50 55 30 60,clip,scale=0.35]{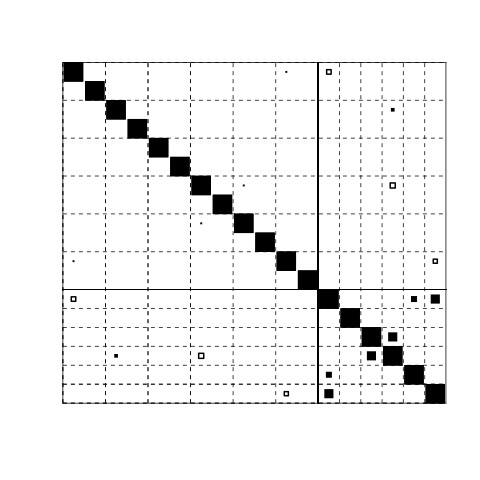}}}
	{\subfloat[Second behaviour]{\includegraphics[trim=50 55 30 60,clip,scale=0.35]{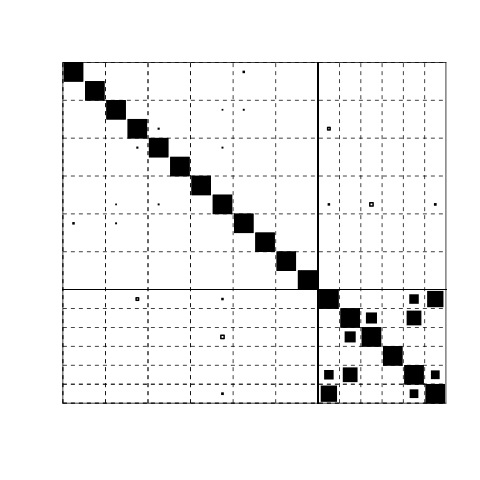}}}\\
	{\subfloat[Third behaviour]{\includegraphics[trim=50 55 30 60,clip,scale=0.35]{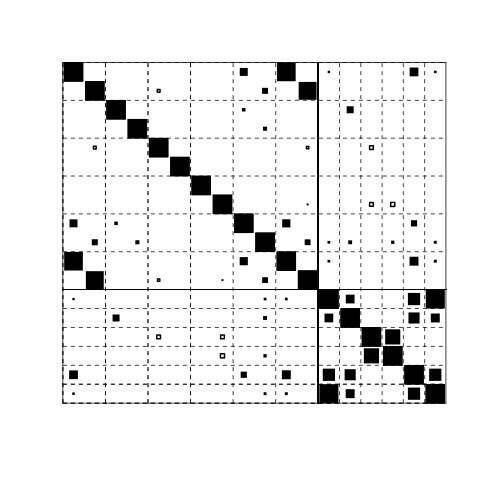}}}
	\caption{Graphical representation of the posterior means of the  correlation matrices  $\boldsymbol{\Omega}_k$s. A filled square indicates a positive value while an empty one  a negative; a square is depicted only if the associated correlation is significantly different from 0. The dimension of the square is proportional to the absolute value of the associated mean correlation coefficient. 
		The  full lines separate the  correlation matrix of the cosine and sine of the circular variables (top-left), the correlation matrix of the linear ones (bottom-right) and the correlation matrix between the linear variables and the sine and cosine of the circular ones (top-right).
		$\hat{K}=3$.
	} \label{fig:Cor}
\end{figure}

In this Section we apply our proposal  on a real dataset,  taken from the movebank website \citep{vanbommel20142}.

\subsection{Data description} Data on   free-ranging Maremma sheepdogs positions are recorded by tracking collars  every 30 minutes.  The behaviour of the dogs is  unknown because there is minimal supervision by their owners and the animals are allowed to  range freely. The dataset was  first analyzed in \cite{vanbommel2014}  with the aim to understand how much space the dogs utilize and  the portion of time that the dogs spent with livestock. Even if the primary purpose was not to identify behavioural modes, \cite{vanbommel2014} results show that the dogs can be clustered in  two states,  one characterized by low speeds and tortuous path at the core of their home ranges (we call   it state  VB1), when they are resting or attending livestock,   and  large step-lengths (i.e. high movement speeds) in  relatively straight lines, related to boundary patrolling or seeing off predators (we call   it state VB2), at the edge of their home ranges.

We   characterize the hidden behaviours by analyzing the turning-angles and the logarithm of the step-lengths\footnote{The logarithm is needed since  the linear components of the JPSN  must be defined over $\mathbb{R}$.} (log-step-lengths) for each dog taken into consideration. We model movement metrics belonging to dogs sharing the same property and observed on a  common time period.
We select  the data from the  ``Heatherlie'' property where between  the 08/02/2012 5:30 and   10/03/2012 17:00, six dogs are observed, having then a time series of 3000 points with 6 circular and 6 linear variables. The  six dogs  have respectively 
107, 63, 231, 43, 31 and 63 missing circular observations and
95, 53, 117, 32, 22 and 45  linear ones. The density estimates of the turning-angles and log-step-lengths can be seen in Figures \ref{fig:Dcirc} and \ref{fig:Dlin}. In the selected property and observational interval,  \cite{vanbommel2014} note that four of the six dogs form one social group responsible for protecting all livestock,  one dog is old   and mostly solitary and the last one  suffers of an extreme social exclusion 
which severely restrict its movements.
The animals that are part of the
social group are often found together, but  regularly they
split into sub-groups.

%
%

\subsection{Results} \label{sec:ress}

\begin{table}[t!]
	\centering
	\begin{tabular}{c|ccc}
		\hline \hline
		& 1&2&3 \\
		\hline
		1& 0.711 &0.181 & 0.108\\
		& (0.679 0.741)& (0.154 0.209)& (0.085 0.132)\\
		2& 0.141 & 0.672 & 0.187\\
		& (0.121 0.164)& (0.640 0.704)& (0.161 0.215)\\
		3&  0.083 & 0.209& 0.708\\
		& (0.065 0.103)& (0.180 0.239)& (0.676 0.739)\\
		\hline
		\hline
	\end{tabular}
	\caption{Posterior mean estimates and 95 \% credible intervals for the transition probability matrix: $\hat{K}=3$.  } \label{tab:simreal}
\end{table}

The model is estimated considering 400000 iterations, burnin 300000, thin 20 and by taking  5000 samples for inferential purposes. As  prior distributions  we choose  $\boldsymbol{\mu}_k,\boldsymbol{\Sigma}_k \sim NIW \left( \mathbf{0}_{6}, 0.001, 25, \mathbf{I}_{6}  \right)$ and $\boldsymbol{\lambda}_k \sim N_{2}\left( \mathbf{0}_2, 50 \mathbf{I}_2 \right)$ that are standard weak informative distributions.  
For the parameter $\rho$, that governs the self-transition probabilities, we decide to use $\rho \sim B(1,1)$, that is equivalent to a uniform distribution over  $[0,1]$, while $\tau \sim G(1, 0.01)$ and  $\gamma \sim G(1, 0.01)$.   The priors of $\rho$, $\tau$ and $\gamma$ induce a prior over $K$ \citep{fox2011} that we evaluated  through simulation,  by using the \emph{degree $\ell$  weak limit approximation} \citep{Ishwaran2002}  with $\ell=1000$;  we found that  ${K}  \in [4,465]$ with a coverage of 95\%.

The model estimates  3  behavioural modes with  $P(K=3|\boldsymbol{\theta}, \mathbf{y})=1$. 
From the analysis of the posterior marginal distributions, Figures    \ref{fig:DestDir} and \ref{fig:DestLin}, and the correlation matrices  $\boldsymbol{\Omega}_k$, Figure \ref{fig:Cor}, we can easily interpret the behaviour modes, we can confirm what \cite{vanbommel2014} found, we  find  connections between our estimated behavioural modes and the states VB1 and VB2 hypothesized by \cite{vanbommel2014}, and we add some new results.  
There are three groups of dogs that share similar marginal distributions: the dogs group one (DG1) composed by the dogs three and four, Figure \ref{fig:DestDir} (c) and (d) and Figure \ref{fig:DestLin} (c) and (d), the dog group two (DG2) is composed by the dogs two and five, Figure \ref{fig:DestDir} (b) and (e) and Figure \ref{fig:DestLin} (d)  and (e), and the dogs one and six  form the third group (DG3), Figure \ref{fig:DestDir} (a) and (f) and Figure \ref{fig:DestLin} (a) and (f).

In the first behaviour, all the dogs are in state VB1. They have small log-step-lengths and there are few movements in a straight line, i.e. the circular densities have low values  at 0. 
%
%
There are few correlations that differ significantly from 0\footnote{A correlation differs significantly from 0 if  its 95\% credible interval (CI) does not contain the 0.}, see Figure \ref{fig:Cor} (a), and they have all mean posterior values (PEs) below 0.5. The stronger correlation (PE 0.474) is between the linear variables of the first and sixth dogs, i.e. the one in the  DG3.

In the second behaviour,  the linear distributions of the dogs in the DG3 have more mass of probability at higher values, but still having mass at low ones, and the correlation between  the log-step-lengths increases (PE  0.840).  The dogs in the  DG3 have  more movements in a straight line. The linear and circular distributions of the dogs in the DG1 and DG2 are  similar to the ones of the first behaviour. The log-step-length of the dog two is correlated with the one of the dog three (PE 0.568) and five (PE 0.757) and the linear variables of the dogs five and six are correlated with PE 0.450. In this behaviour the dogs in the DG3 move to the state VB2 while the ones of the DG1  and DG2 remain  in the state VB1.

In the third behaviour the linear distributions of the dogs in the DG3 put all the mass of probability on high values and the circular distributions have  two modes, with more or less the same heights, at about 0 and $3.141$. The correlation between the linear variables of the dogs in the DG3 increases, with respect to the second behaviour, and it is almost one  (PE   0.982). The two dogs change direction one  accordingly to the other, since the cosine and sine of the circular variables are highly correlated (PEs 0.958 and 0.916).
The linear distributions of the dogs in the  DG2 move their mass of probability to higher values and the distributions resemble the ones of the dogs in the DG3, second behaviour.  The circular distributions of the dogs in the  DG2  are close to the circular uniform.  
The circular and linear distributions of the dogs in the DG1 are similar to the one of the first and second behaviours but  their linear variables are now correlated (PE 0.781).  In this behaviour the dogs in the DG3  remain  in the state VB2  but, with respect to the second behaviour, they increase the amount of  movement (in terms of  higher step-length and more tortuous path). The dogs in the DG2 are in the state VB2 while the one in the DG1  remain in the state VB1.   The dogs in the DG2 and DG3 have all the  linear variables correlated. 

The fourth dog is the one that suffers of an extreme social exclusion since its variables (circular and linear) are never  correlated with the ones of the other dogs, with the exception of the third. The third dog  is, probably, the  old one since it does not move a lot, see Figure \ref{fig:DestLin} (c), it is solitary, i.e.  it bonds (in terms of correlation) only with  the socially excluded one and, occasionally, with the dog two (second behaviour). The dogs in the DG2 and DG3  form one social group in the third behaviour, i.e. all the linear variables are correlated, and they split into subgroup in the behaviours one and two. 

From Table \ref{tab:simreal} we see that  there is a strong self-transition in all three behaviours (respectively PE 0.711, 0.672 and 0.708). 
The CIs of the probabilities to move to a new empty behaviour $\left(\sum_{k=4}^{\infty}\pi_{jk}, \, j=1,2,3\right)$, not shown in Table \ref{tab:simreal}, have always right side limit below 0.00001.

\subsection{Comparisons with other emission distributions} \label{sec:conf}

\begin{table}[t!]
	\centering
	\begin{tabular}{c|ccccc}
		\hline \hline
		&JPSN& VMLG& VMLW& WCLG & WCLW \\ \hline
		$\hat{K}$ 	& 	3  &   14  &  14   &14      &  13   \\
		$CRPSc$ 	& 	0.489  &	0.491 & 0.499 & 0.492 &   0.501   \\
		$CRPSl$ 	& 2.342	  &  2.782	  & 3.225 &  2.466	&   2.965   \\\hline \hline
	\end{tabular}
	\caption{Estimated number of non-empty behaviours ($\hat{K}$), mean CRPS for circular ($CRPSc$) and linear ($CRPSl$) variables, for the five models based on the JPSN, VMLG, VMLW, WCLG and WCLW.} \label{tab:mod5}
\end{table}
\begin{figure}[t!]
	\centering
	{\subfloat[]{\includegraphics[scale=0.3]{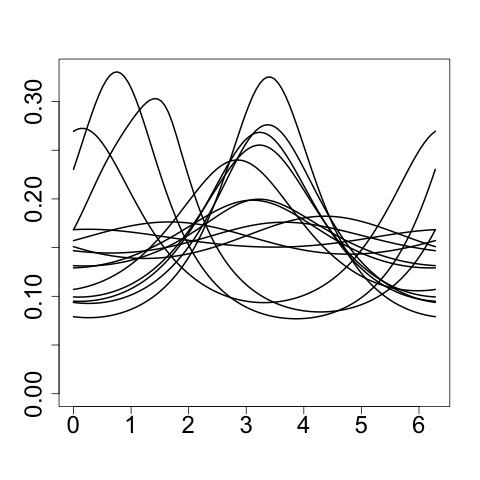}}}
	{\subfloat[]{\includegraphics[scale=0.3]{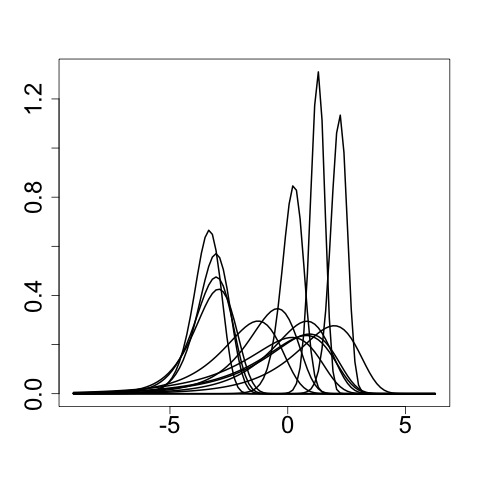}}}
	\caption{Posterior marginal densities of the turning-angle (a) and the log-step-length (b) of the first dog estimated with the emission distribution WCLG.
	} \label{fig:DestEX}
\end{figure}
In this Section we  show that our proposed emission distribution performs better, on the data we used in Section \ref{sec:ress}, than the standard distributions used in the literature. We are going to estimate sHDP-HMMs with different emission distributions and we compare the results.  Unfortunately there are not  measures of goodness of fit, such as the AIC or the BIC, when the  model is  based on the  sHDP.  Then we decide to base our  comparison between models in terms of missing observations estimate, i.e. predictive ability,  and behavioural modes interpretability.

To have a measure of how the model  estimates the missing, we randomly select, for each circular and linear variables, 10\% of the observations.  We treat them as missing and,  using the \emph{continuous ranked probability score} (CRPS) \citep{Matheson1976},   we compare the holdout values with the associated posterior distributions. The CRPS is a proper scoring rule that can be easily computed for both circular \citep{grimit2006} and linear  \citep{Gneiting2007} variables using the MCMC output. Let $\mathcal{C}_i$ be the set of time points where the $i^{th}$ value of the circular variable is setted as missing, $\mathcal{L}_{j}$ be the ones of the $j^{th}$ linear variable and let $\theta_{ti}^b$, $t \in \mathcal{C}_i$ and $y_{tj}^b$, $t \in \mathcal{L}_j$ be respectively the $b^{th}$ posterior sample of $\theta_{ti}$ and $y_{tj}$.  A Monte Carlo approximation of the CRPS for a circular variable is computed as
\begin{equation}
	CRPSc_{i} \approx \frac{1}{B}\sum_{b=1}^Bd(\theta_{ti},\theta_{ti}^b)-\frac{1}{2 B^2} \sum_{b=1}^B \sum_{b^{\prime}=1}^B d(\theta_{ti}^b,\theta_{ti}^{b^{\prime}}), \, t \in \mathcal{C}_i,
\end{equation}
where $d(\cdot,\cdot)$ is the angular distance. The CRPS for a linear variable is approximated with 
\begin{equation}
	CRPSl_{j} \approx \frac{1}{B}\sum_{b=1}^B |y_{tj}-y_{tj}^b|-\frac{1}{2 B^2} \sum_{b=1}^B \sum_{b^{\prime}=1}^B |y_{tj}^b-y_{tj}^{b^{\prime}}|, \, t \in \mathcal{L}_j.
\end{equation}
We then compute the overall mean CRPS for the circular variables, $CRPSc= \frac{1}{n}\sum_{i=1}^n CRPSc_{i}$, and the linear ones, $CRPSl= \frac{1}{q}\sum_{j=1}^q CRPSl_{j}$, and we use these two indeces to measure the ability of the model in estimating the missing observations.

It is 
generally  supposed, in the literature,  that the turning-angle is distributed as a    von Mises   \citep{Langrock2012,Holzmann2006,Eckert2008} or a wrapped Cauchy \citep{Langrock2012,Eckert2008,morales2004,Holzmann2006} while the gamma \citep{Langrock2012,Holzmann2006} or the Weibull \citep{Langrock2012,morales2004} are used for the step-length; these distributions are compared with our proposal.  In the model specification, Section \ref{sec:HMM}, we assume that each  linear variable belongs to $\mathbb{R}$ and then, instead of the gamma and Weibull, we use  the log-gamma and log-Weibull, i.e. the distributions that arise by taking the log of, respectively, a random variable  gamma or Weibull distributed.  The model in Section \ref{sec:HPD-HMM}  is compared with the ones based on the von Mises and the log-gamma (VMLG), the von Mises and the log-Weibull (VMLW), the wrapped Cauchy and the log-gamma (WCLG), the wrapped Cauchy and the log-Weibull (WCLW). There is not an obvious way to introduce dependence between the movement metrics on the model VMLG, VMLW, WCLG and WCLW, and, in the literature,    they are generally supposed to be independent (see for example \cite{morales2004} or \cite{Langrock2012}). Then, in these models, we assume the following:
\begin{equation}
	f(\boldsymbol{\theta}, \mathbf{y}|\{z_t\}_{t \in \mathcal{T}}  \{\boldsymbol{\psi}_{k}\}_{k \in \mathcal{K}})  =   \prod_{t \in \mathcal{T}} \prod_{k \in \mathcal{K}} \left[ \prod_{i=1}^n         f(\theta_{ti}|\boldsymbol{\psi}_{z_t}  )\prod_{j=1}^q   f(y_{tj}|\boldsymbol{\psi}_{z_t}  ) \right]^{I(z_t,k)}. 
\end{equation}
We use a $G(1,0.5)$ as prior for the \emph{shape} and \emph{rate} parameters of the log-gamma and the log-Weibull, that is a standard weak-informative prior. For the 
two parameters of the wrapped Cauchy, one defined over $[0, 2 \pi)$ and one over $[0,1]$, we use uniform distributions in the respective domains while on the two parameters of the von Mises, one defined over $[0,2 \pi)$ and one over $\mathbb{R}^+$, we use respectively the non-informative $U(0, 2 \pi)$ and the  weak informative $G(1,0.5)$.  As prior distributions for the sHDP parameters, we use the same used in Section \ref{sec:real}. 

In Table \ref{tab:mod5} we can see the  estimated number of non-empty behaviours $(\hat{K})$,   $CRPSc$s and $CRPSl$s. The predictive ability of our model outperforms all the others in both CRPS for circular and linear variables.  The  models based on the VMLG, VMLW, WCLG and WCLW estimate a larger number of behaviours, with respect to our proposal, i.e. in three of them  $\hat{K}=$14   and in one  $\hat{K}=$13; we can see an example of the estimated behaviours in Figure \ref{fig:DestEX}.  It is challenging to give an interpretation to these  behaviours and moreover such a large number  of behaviours does not increase the predictive  ability of the models, see Table \ref{tab:mod5}.

A possible explanation of  why the models based on the  VMLG, VMLW, WCLG and WCLW estimate 13 or 14 behaviours can be found in  \cite{mastrantonio2015}. 
They simulate  datasets using  bivariate emission distributions  with dependent components,  bimodal marginals  for the circular variable,  
with the aim  to understand what happens if, on the  simulated datasets, are estimated  HMMs with emission distributions that do not allow for dependent components and bimodality in the circular marginal. They found that the number of behaviours is generally overestimated, since each behaviour is separated into two, one for each mode of the circular distribution. 		\\
In our real data application, we have six circular and six linear variables,  some of them are correlated and often  the marginal circular distributions have  two modes, see Figure \ref{fig:DestDir}. 
If we assume independence between the circular and linear variables, as we did in the model base on the  VMLG, VMLW, WCLG and WCLW, and the marginal circular distributions are unimodal, as  the von Mises and the wrapped Cauchy,  then we can expect   a large  number of behaviours.

\section{Conclusions} \label{sec:conc}

The primary objective of this work, motivated by our real data, was to introduce an HMM capable of modelling a group of animals, taking into account possible correlations between the associated  movement metrics. 
For this reason we introduced a new multivariate circular-linear distribution, namely the JPSN. The new distribution has  dependent components and it  is used as emission distribution in the HMM. The HMM  was estimated under a non-parametric Bayesian framework and we showed how to implement the MCMC algorithm. 

The model, applied to the real data example, confirmed   known results and  added new ones.
We  showed that our emission distribution outperforms the most used in the literature in terms of predictive ability and the estimated behaviours are more easily interpretable.

Future work will lead us to incorporate covariates  to  model the circular-linear mean and variance of the  JPSN and we will  explore  different temporal dependence structures, such as the semi-HMM or the autoregressive-HMM.

\bibliographystyle{natbib}
\bibliography{all}
\end{document}